\documentclass[
 preprint,
superscriptaddress,
bibnotes,
showkeys,
amsmath,amssymb,
aps,prb,
floatfix,
]{revtex4-1}

\usepackage{graphicx}% Include figure files
\usepackage{dcolumn}% Align table columns on decimal point
\usepackage{bm}% bold math
\usepackage{natbib}
\usepackage{color}
\begin{document}

\preprint{APS/123-QED}
\title{Micron-scale measurements of low anisotropic strain response of local $T_c$ in Sr$_2$RuO$_4$}

\author{Christopher A. Watson}
\affiliation{
 Stanford Institute for Materials and Energy Sciences,
SLAC National Accelerator Laboratory,\\
2575 Sand Hill Road, Menlo Park, CA 94025, USA
}

\author{Alexandra S. Gibbs}
\altaffiliation[Current address: ]{ISIS Neutron and Muon Source, STFC Rutherford Appleton Laboratory, Harwell Campus, Didcot, OX11 0QX, UK}
\affiliation{Scottish Universities Physics Alliance, School of Physics and Astronomy, University of St. Andrews, St. Andrews KY16 9SS, United Kingdom}
\author{Andrew P. Mackenzie}
\affiliation{Scottish Universities Physics Alliance, School of Physics and Astronomy,
University of St. Andrews, St. Andrews KY16 9SS, United Kingdom}
\affiliation{Max Planck Institute for Chemical Physics of Solids, N\"othnitzer Stra{\ss}e 40, Dresden 01187, Germany}
\author{Clifford W. Hicks}
\affiliation{Max Planck Institute for Chemical Physics of Solids, N\"othnitzer Stra{\ss}e 40, Dresden 01187, Germany}
\author{Kathryn A. Moler}
\affiliation{
 Stanford Institute for Materials and Energy Sciences,
SLAC National Accelerator Laboratory,\\
2575 Sand Hill Road, Menlo Park, CA 94025, USA
}

\date{\today}

\begin{abstract}
Strontium ruthenate (Sr$_2$RuO$_4$) is a multiband superconductor that displays evidence of topological superconductivity, although a model of the order parameter that is consistent with all experiments remains elusive. We integrated a piezoelectric-based strain apparatus with a scanning superconducting quantum interference device (SQUID) microscope to map the diamagnetic response of single-crystal Sr$_2$RuO$_4$ as a function of temperature, uniaxial pressure, and position with micron-scale spatial resolution. We thereby obtained local measurements of the superconducting transition temperature $T_c$ vs. anisotropic strain $\epsilon$ with sufficient sensitivity for comparison to theoretical models that assume a uniform $p_x\pm ip_y$ order parameter. We found that $T_c$ varies with position and that the locally measured $T_c$ vs. $\epsilon$ curves are quadratic ($T_c\propto\epsilon^2$), as allowed by the C$_4$ symmetry of the crystal lattice. We did not observe the low-strain linear cusp ($T_c\propto \left| \epsilon \right|$) that would be expected for a two-component order parameter such as $p_x\pm ip_y$. These results provide new input for models of the order parameter of Sr$_2$RuO$_4$. 

\end{abstract}

\maketitle

\section{\label{sec:Intro}Introduction}

Since its discovery in 1994,\cite{Maeno_1stpaper} strontium ruthenate (Sr$_2$RuO$_4$) has generated considerable theoretical and experimental interest as a candidate topological superconductor.\cite{Mackenzie_evenodder, Kallin_review, Maeno_tripletreview}  It was proposed shortly after its discovery that the pairing in Sr$_2$RuO$_4$ might be spin triplet, with an orbital component that is chiral with irreducible representation $p_x \pm ip_y$.\cite{Sigrist_rice} The spin part of the order parameter has been probed in multiple NMR experiments, the results of which are consistent with expectations for spin triplet pairing.\cite{NMRNQRReview} The hypothesis of time-reversal-symmetry breaking and hence chiral orbital order is supported by muon spin rotation ($\mu$SR),~\cite{Luke_muSR}  polar Kerr effect,\cite{Xia_Kerr} the critical current of Sr$_2$RuO$_4$/conventional superconductor junctions, \cite{Junction_measurements} and other measurements. However, the inferred sizes of chiral domains vary greatly between those measurements;\cite{Kallin_review} edge currents are expected\cite{Matsumoto_Sigrist} to appear with a chiral order parameter, yet are not observed;\cite{Hicks_SQUID, Curran_Hall} and there are other compelling results that do not follow expectations for chiral order.\cite{Mackenzie_evenodder, Kallin_review, Maeno_tripletreview}  Recent theoretical analysis suggested that the predicted edge current magnitude may be substantially smaller than originally suggested.\cite{Huang_nocurrents, *Lederer_nocurrents, *Scaffidi_nocurrents} Overall, the order parameter of Sr$_2$RuO$_4$ remains an important and intriguing question.

\begin{figure}
\includegraphics[width=.5\columnwidth]{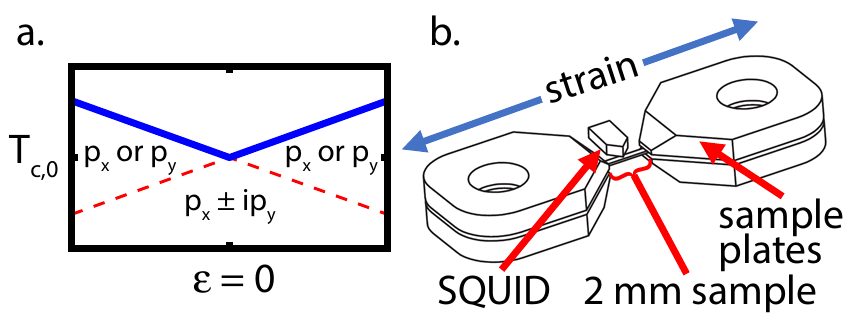}
\caption{\label{strainjig}Integration of the strain apparatus, similar to that described previously,\cite{Hicks_strain} with scanned probe microscopy to test for a chiral order parameter. (a) Our hypothesis: the dependence of the superconducting transition temperature on anisotropic strain, $T_c(\epsilon)$, is expected to have a linear cusp at zero strain if the order parameter is chiral. (b) Schematic of the sample plates with a mounted sample and the scanning SQUID chip nearby. The SQUID chip is positioned such that its micron-scale pickup loop (not shown here) scans the sample surface. The exposed portion of the sample is $\sim 2$ mm long, allowing the sensor to fit between the sample plates.  Stress is applied along the length of the sample, as indicated.}
\end{figure}

One proposal to test for chiral order involves applying in-plane uniaxial pressure,\cite{Sigrist_pressure} or in-plane magnetic field,\cite{Gorkov} to lift the C$_4$ symmetry of the unstressed lattice and therefore the degeneracy of the $p_x$ and $p_y$ components, resulting in a split transition [Fig.~\ref{strainjig}(a)]. In a measurement that is sensitive mainly to the onset of superconductivity, such as ac susceptibility, the observed superconducting transition temperature $T_c$ will be that of the component with the higher $T_c$.  Therefore, the dependence of $T_c$ on strain $\epsilon$ should have a linear cusp at $\epsilon = 0$, i.e. a term in $T_c(\epsilon)$ that is proportional to $\left|\epsilon\right|$. So far, experimental studies have revealed no evidence of such a split transition under either in-plane uniaxial stress\cite{Hicks_SRO,Steppke_SRO} or in-plane magnetic field.\cite{Yaguchi_field,*Mao_field}  However, due to a strong underlying strain dependence of $T_c$ and the possibility of a cusp being rounded by sample inhomogeneity, the uniaxial stress experiments\cite{Hicks_SRO,Steppke_SRO} did not place tight bounds on the magnitude of an $\left|\epsilon\right|$ term.  The experimental limits are comparable to theoretical estimates for the magnitude of this term for $p_x\pm ip_y$ order;\cite{Steppke_SRO} therefore, the previous measurements do not constitute rigorous tests of this predicted signature of chiral order.  

The samples used in those experiments were of high quality, and further improvement in sample quality might not be practical. Therefore, to measure $T_c(\epsilon)$ with better resolution, we turn to scanned-probe measurements.  We describe here the first successful integration of a piezoelectric-based apparatus for \textit{in situ} application of uniaxial pressure with low-temperature scanned probe microscopy.  Our probe is a scanning superconducting quantum interference device (SQUID) susceptometer,\cite{Kirtley_SQUID} which can be used to measure $T_c$ locally by detecting the onset of Meissner screening.  By obtaining scans of the ac susceptibility as a function of temperature, we demonstrate our ability to resolve spatial inhomogeneity in the sample and find the scale of $T_c$ inhomogeneity to be approximately 50 mK.  By positioning the susceptometer on the surface of the sample in regions with particularly high local homogeneity, we obtain measurements of the diamagnetic susceptibility as a function of temperature and observe that the superconducting transitions are rounded only at the level of 1 mK, implying that $T_c$ within the measurement volume ($\sim 100~\mu$m$^3$) is homogeneous to within at least this level.  With this improved sensitivity, we then measure the low-strain response of $T_c$ through zero strain.  We show that the strain dependence of $T_c$ is in good agreement with a purely quadratic response, placing an upper bound on any $\left|\epsilon\right|$ term that now does impose a meaningful constraint on theory.

\section{\label{sec:Setup}Methods}
With a wire saw, we cut a beam, oriented in the $\langle 100\rangle$ lattice direction, from a rod of Sr$_2$RuO$_4$ grown using the floating zone method.\cite{MaoGrowth} Uniaxial stress applied along the $\langle 100\rangle$ direction has a much stronger effect on $T_c$ than along the $\langle 110\rangle$ direction. We polish the surface to obtain a uniform cross-section and a smooth upper surface for scanning.  

We use a piezoelectric-based strain apparatus similar to that described previously,\cite{Hicks_strain} modified for compatibility with our scanned probe microscope.  In particular, the relatively large dimension of our SQUID chip\cite{Kirtley_SQUID} requires an exposed sample length of at least $2~\text{mm}$, further requiring a larger net displacement to achieve a specified strain.  We accomplish this by using longer piezoelectric actuators (Physik Instrumente P-885.11) and a symmetric design. 

We mount the sample in the strain apparatus between lightly abraded titanium sample plates [Fig.~\ref{strainjig}(b)] using a thermally conducting, electrically insulating epoxy (Epo-Tek H70E), cured according to its lowest-temperature standard curing schedule (80$^\circ$C for 90 minutes).  At low temperatures, we drive the piezoelectric actuators with a high voltage source (Keithley 2410 High Voltage SourceMeter), filtered by a $1~\text{M}\Omega$ resistor.  We determine the strain setting \textit{in situ} as the displacement applied to the sample mounts, measured with an integrated parallel plate capacitive sensor, divided by the effective length of the sample.  We take the effective length to be 2.3 mm, slightly longer than the actual exposed length, to account for deformation within the ends of the sample, as described in more detail in Sec.~\ref{sec:Results} below.

The scanning SQUID susceptometer is of the same design as those that have been previously characterized.\cite{Kirtley_SQUID}  It has a pickup loop with a $1~\mu\text{m}$ inside radius and a concentric, single-turn field coil of $2.5~\mu\text{m}$ inside radius [Fig.~\ref{tempseries}(a)] that allows us to apply a local field. Applying a low-frequency ac current to the field coil and detecting the resultant flux through the pickup loop measures the mutual inductance between the pickup loop and field coil, which is modified by the presence of any magnetic sample. The vacuum response of the SQUID to the applied ac field is canceled by a nominally identical counter-wound pickup loop--field coil pair located far from the sample surface; as a result, a nonzero signal indicates a sample's response to the applied local field.  In the case of superconductivity, the repulsion of the applied field due to the Meissner effect results in a reduced flux near the sample and a negative total mutual inductance.  

The SQUID chip is mounted on a brass foil cantilever approximately 8 mm long, 3 mm wide, and 25 $\mu$m thick. To determine the spatial variation in $T_c$, we raster the susceptometer in a plane parallel to and just above the sample surface with the sample at various temperatures through the bulk $T_c$.  We then select points in regions with highly homogeneous $T_c$ for more careful study.  To precisely measure $T_c$ at a point, we place the SQUID chip in light mechanical contact with the sample, enough to deflect the cantilever by a few hundred nanometers, to ensure a constant position and sensor-sample separation. We control the sample temperature by digitally switching a heater between high and low settings, choosing voltages and dwell times  to reduce thermal hysteresis to below 1 mK while capturing the full range of $T_c$ values observed in the strain series.

\section{\label{sec:Results}Results}

\begin{figure*}
\includegraphics[width=\linewidth]{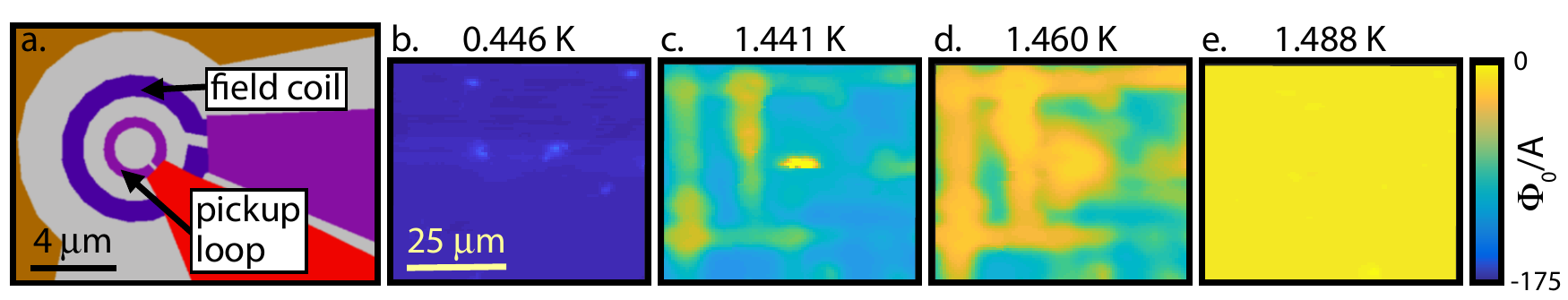}
\caption{\label{tempseries}Temperature series of susceptibility scans demonstrating spatial inhomogeneity of the superconducting transition. (a) A computer-aided drawing of the SQUID pickup loop-field coil pair, realizing a micron-scale ac susceptometer. (b) At base temperature (446 mK), the sample is strongly superconducting as demonstrated by a large, nearly uniform diamagnetic susceptibility, while (e) at the highest temperature shown (1.488 K), the sample has no magnetic susceptibility as measured at the scan height.  (c, d) At intermediate temperatures, linear features where $T_c$ is lower become visible.}
\end{figure*}

An example of a temperature series of susceptibility scans is shown in Fig.~\ref{tempseries}.  In Fig.~\ref{tempseries}(b), where $T$ = 0.446 K, the sample is strongly diamagnetic\footnote{It is reduced as compared to the bare mutual inductance of the actual pickup loop-field coil pair by the finite penetration depth as well as a geometric factor which depends on the sensor height and angle.} and the diamagnetism is highly homogeneous, consistent with being deep in the Meissner state.  At temperatures near the bulk $T_c$ [Fig.~\ref{tempseries}(c) and (d)], the diamagnetism shows stronger inhomogeneity.  There are linear features where $T_c$ is locally reduced, and overall $T_c$ is inhomogeneous on a 10 $\mu$m length scale. By 1.488 K [Fig.~\ref{tempseries}(e)], there is no detectable susceptibility signal at the scan height in this region of the sample.\footnote{In the analysis below, we define $T_c$ more strictly as the onset of diamagnetic susceptibility as measured with the SQUID in contact.  Because the sensitivity in this configuration is somewhat higher than while scanning, the reported $T_c$'s are in some cases higher than the temperature at which the sample appears to be normal in the scans.}  These scans, with scans at additional temperatures, show that the scale of $T_c$ inhomogeneity over this portion of the sample is $\sim 50$ mK. 

\begin{figure}
\includegraphics[width=.5\columnwidth]{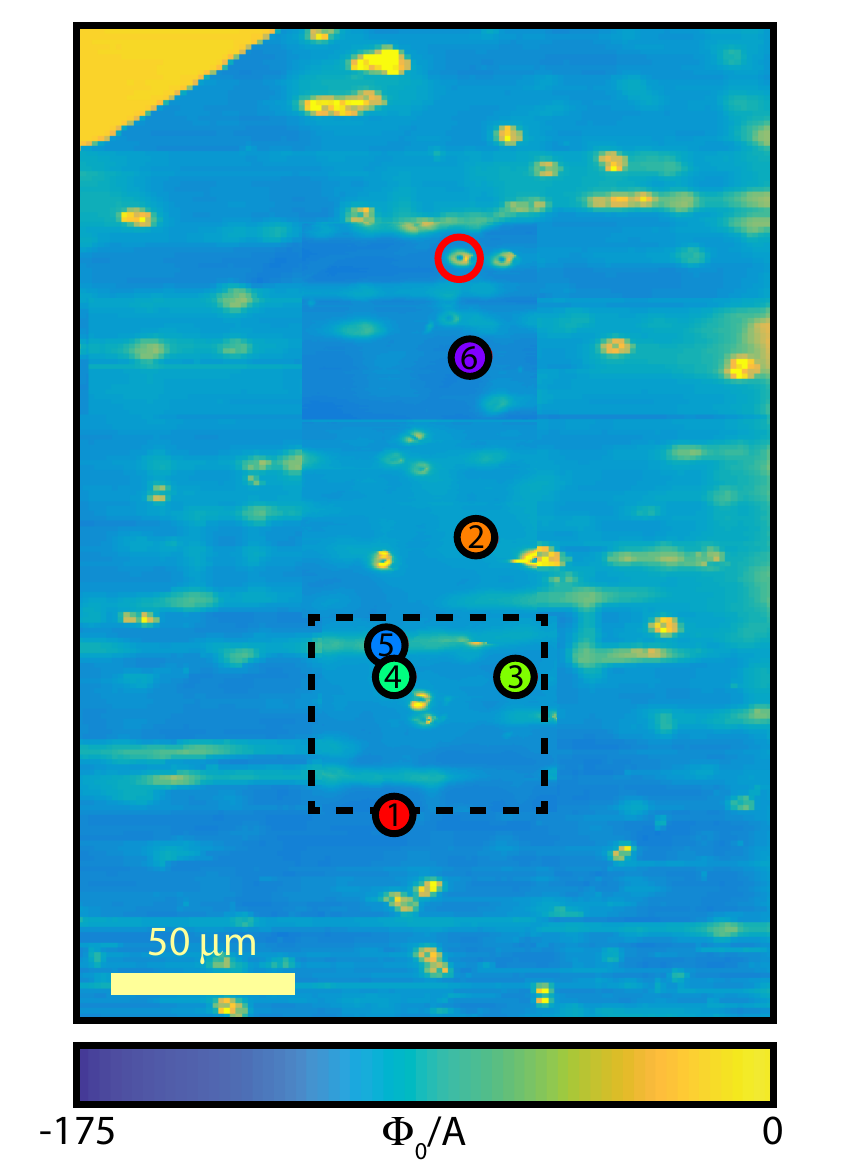}
\caption{\label{mosaic}Mosaic of susceptibility scans taken at 1.43 K, just below the bulk transition temperature.  Markers indicate the nominal positions where $T_c(\epsilon)$ was measured; these data are shown in Fig.~\ref{allstraincurves}.  The dashed black box indicates the location of Fig.~\ref{tempseries}.  The yellow feature at top left is due to the SQUID touching the sample clamp.  Small ringlike features seen throughout, such as the one circled in red at top center, are an imaging artifact due to superconducting vortices.\cite{[{The artifact is due to a temporary loss of sensitivity when scanning over a large, discrete dc magnetic signal (such as that seen for a vortex) with insufficient gain in the flux feedback, as previously described in Appendix A of }] Eric_thesis}}
\end{figure}

Figure~\ref{mosaic} shows a mosaic of susceptibility scans at 1.43 K covering approximately $70000~\mu\text{m}^2$.  Inhomogeneity in $T_c$ is visible across this area. One of the sample clamps is visible in the mosaic, at top left, while several superconducting vortices can also be seen as ring-like imaging artifacts (such as the one circled in red at top center).\cite{[{The artifact may be due to actual motion of the vortices under the applied ac field, or could be due to a temporary loss of sensitivity when scanning over a large, discrete dc magnetic signal (such as that seen for a vortex) with insufficient gain in the flux feedback, as previously described in Appendix A of }] Eric_thesis}  To measure the transition temperature as a function of applied strain, we choose several locations on the sample as indicated by the numbered markers.  These points are separated from prominent inhomogeneity by at least $10~\mu\text{m}$, except for point 5 which was deliberately chosen on a linear feature.\footnote{Based on the measured minimum $T_c$, it appears that at least Points 1 and 7 were also near such features, though this was not intentional and reflects a deviation between the nominal location and the actual measurement location.  Sweeps 7 were taken at the same nominal location as 6.}  

\begin{figure}
\includegraphics[width=.5\columnwidth]{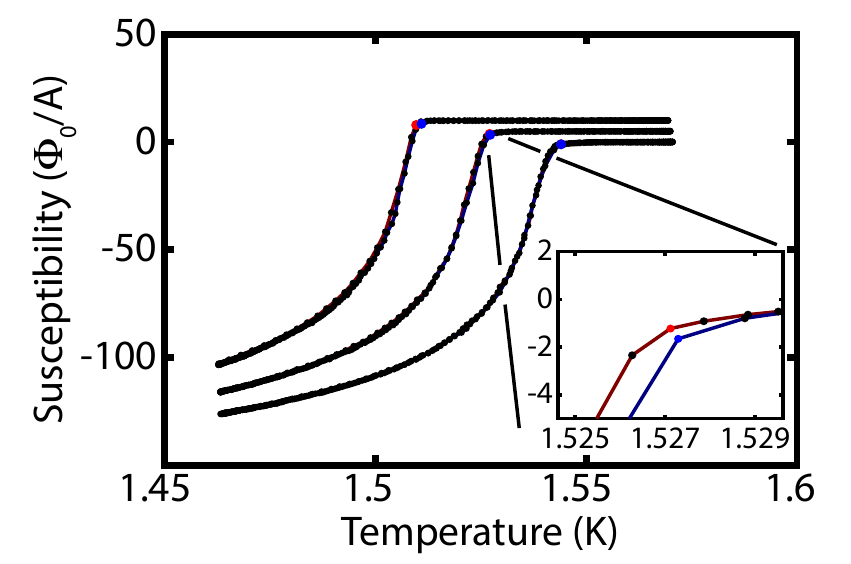}
\caption{\label{transition}Magnetic susceptibility as a function of temperature for three values of applied displacement ($-0.50$, $-1.59$, and $-2.01$ $\mu$m from left to right), at Point 1 in Fig.~\ref{mosaic} (offset for clarity).  Both warming (black dots connected by dark red lines) and cooling (black dots with dark blue lines) sweeps are plotted, demonstrating the small thermal hysteresis, while the sharp, linear onset of diamagnetic susceptibility indicates high sample quality, as discussed in the main text.  The critical temperatures are defined by a threshold susceptibility of $-1~\Phi_0/$A and are indicated by red (warming) and blue (cooling) dots.  Inset shows an enlargement of the transition point for the middle trace.}
\end{figure}

At each location, we place the SQUID sensor in contact with the sample and sweep the temperature back and forth through the transition.  We collect susceptibility data continuously and synchronously with temperature data. Figure~\ref{transition} shows the resulting plots for three different values of strain for point 1.  The most striking feature of these susceptibility traces is the sharpness of the onset of measurable diamagnetic susceptibility:  the transition at $T_c$ is rounded by less than 1 mK, in contrast with the 50 mK large-scale spatial inhomogeneity.  That is, while the sample has large-scale inhomogeneity, $T_c$ is generally homogeneous to better than 1 mK over the approximately 100 $\mu$m$^3$ volume measured by the susceptometer.  It is also noteworthy that the susceptibility varies linearly with $T$ just below $T_c$. As described previously,\cite{Kirtley_paramagnetic_superconductor} in the case of weak, bulk superconductivity, as expected for a 3D superconductor just below $T_c$, the magnetic susceptibility as probed by the susceptometer is proportional to $\lambda^{-2}$.  Sufficiently close to $T_c$, where the penetration depth exceeds the Pippard coherence length, the superconductor is always in the local (London) limit, and the temperature dependence of the penetration depth is given by $(T-T_c)^{-1/2}$, yielding the observed 
linear behavior.\cite{Tinkham}  This mean-field behavior of the superconducting transition in Sr$_2$RuO$_4$ results from its low $T_c$ and relatively long coherence length of 75 nm (atypical for an unconventional superconductor\cite{Mackenzie_evenodder}), and implies that the sample is locally of high quality and that the effects of fluctuations are modest.  

We observe no consistent, systematic variation of the shape of the susceptibility versus temperature curves with applied strain over our applied strain range.  For each curve, we take $T_c$ to be the onset of measurable diamagnetic susceptibility, with a threshold of $-1~\Phi_0/$A.  (We tested a variety of thresholds from $-0.5$ to $-20~\Phi_0/$A and found that the choice of threshold did not qualitatively affect our conclusions.)  In the bulk, weak superconducting limit, this threshold corresponds to a penetration depth of $20 \pm 10~\mu$m, where the error bar results from uncertainty in the sensor-sample separation.\cite{Kirtley_paramagnetic_superconductor}  In the following discussion, we average the $T_c$'s determined from the warming and cooling traces together.  

We now turn to the strain dependence of $T_c$.  As described above, we infer the applied strain from a measurement of an integrated parallel plate capacitive sensor.  Accurate determination of the strain therefore requires characterization of a parasitic, parallel capacitance from the cryostat wiring, the effective length of sample, and the applied displacement at which the zero strain condition is locally achieved.  With the piezoelectric stacks grounded and a known capacitor spacing, the offset capacitance is extracted by measuring the capacitive sensor on the table top and as installed in the cryostat.  Because the cryostat wiring is comprised of twisted pairs, the parasitic capacitance is not fixed between cooldowns and therefore cannot be exactly compensated.  We take the strained length of the sample to be 2.3 mm, slightly longer than the actual exposed length, to account for the fact that the strain relaxes within the epoxy joins over a nonzero distance.  An error in the effective length would correspond to a small overall stretch of the strain axis, but should not substantially alter the strain dependence.

\begin{figure}
\includegraphics[width=.5\columnwidth]{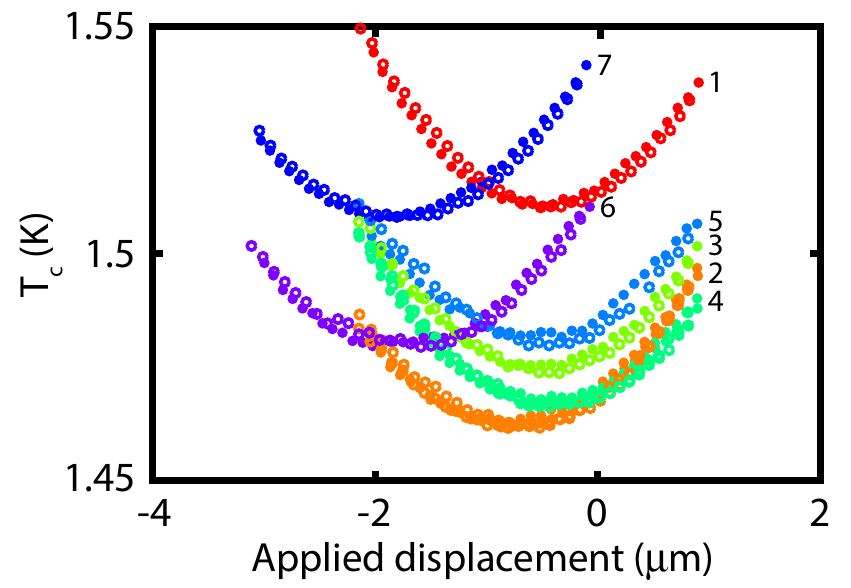}
\caption{\label{allstraincurves}Superconducting transition temperature as a function of applied displacement, measured locally in several locations as indicated in Fig.~\ref{mosaic}.  1 $\mu$m applied displacement corresponds to an estimated strain of $0.043\%$.  Filled and open circles indicate increasing (compressive to tensile) and decreasing (tensile to compressive) sweep directions, respectively, demonstrating hysteresis due to slipping at the epoxy mount.  Horizontal offsets in the data reflect the unknown displacement of the sample mount corresponding to a local zero strain condition}
\end{figure}

The local measurements of $T_c$ versus applied displacement at the six selected points, as well as one additional point of uncertain location, are shown in Fig.~\ref{allstraincurves}; 1 $\mu$m of applied displacement corresponds to a strain of approxiately 0.043\%.  At each point, the dependence is essentially quadratic, independent of the local minimum $T_c$.  There is slight hysteresis in the measured $T_c$ versus applied displacement, which we attribute to slipping of the sample within the epoxy, which turned out not to bond strongly to the sample.  This slipping becomes very clear at large applied displacements, where $T_c$ is observed to saturate, in contrast with previous bulk measurements which show $T_c$ continuing to increase strongly.\cite{Hicks_strain,Steppke_SRO,Barber_SRO}  In Fig.~\ref{allstraincurves}, we show only the low-strain data where $T_c$ is a well-behaved function of displacement and hysteresis is small.  The offsets between the curves along the applied displacement axis indicate the extent to which the sample slipped in the epoxy from one run to the next.

None of the curves in Fig.~\ref{allstraincurves} has an obvious cusp. As the strain range of each curve is limited, however, we must consider whether these curves definitely cross zero strain. Under uniaxial pressure, the strain tensor contains both a component of $B_{1g}$ symmetry ($\epsilon_{xx} = -\epsilon_{yy}$) and components of $A_{1g}$ symmetry ($\epsilon_{xx} = \epsilon_{yy}$, $\epsilon_{zz}\neq 0$).\cite{Palmstrom_Ba122} The latter components add a linear term $T_c \propto \epsilon$ to the strain dependence of $T_c$, and if the coefficient of this term is larger than that of the possible cusp term ($T_c \propto |\epsilon|$), the minimum transition temperature $T_{c,min}$ will not occur at zero strain.

The simplest argument that these curves cross zero strain is that the sample slipped in the epoxy on both the compression and tension ends of these curves. We can also consider more carefully the strain at which the minimum in $T_c$ is expected to occur in the absence of a strong cusp term. We compare the magnitudes of the quadratic and linear terms from bulk data (that is, $a$ and $b$ in $T_c(\epsilon) = a\epsilon^2+b\epsilon+c$). Previous bulk measurements of ac susceptibility with strain applied along the $\langle 100\rangle$ direction yielded $a \sim 6$ K/\%$^2$,\cite{Hicks_SRO,Steppke_SRO,Barber_SRO} and measurements of the jump in ultrasound velocity at the superconducting transition yielded $\sim 5-7$ K/\%$^2$.\cite{Matsui_ultrasound, *Matsui_ultrasound2}  When stress was applied along a $\langle 110\rangle$ direction, on the other hand, the quadratic term was much weaker and a linear coefficient of $b \sim 125$ mK/\% was measured.\cite{Hicks_strain}  The elastic moduli of Sr$_2$RuO$_4$ do not have strong in-plane anisotropy,\cite{PaglionePRB02} so pressure along $\langle 100\rangle$ and $\langle 110\rangle$ will yield similar $\epsilon_{xx} + \epsilon_{yy}$ and $\epsilon_{zz}$ strains.  We can therefore expect similar linear coefficients for the two pressure axes, meaning that  $T_{c,min}$ should occur at a strain of $\epsilon \sim -b/2a \sim -0.01\%$.  The curves in Fig.~\ref{allstraincurves} each span a strain range exceeding 0.1\%, meaning that $\epsilon = 0$ is within the plotted ranges and very close to the minimum of these curves.

\begin{figure*}
\includegraphics[width=\linewidth]{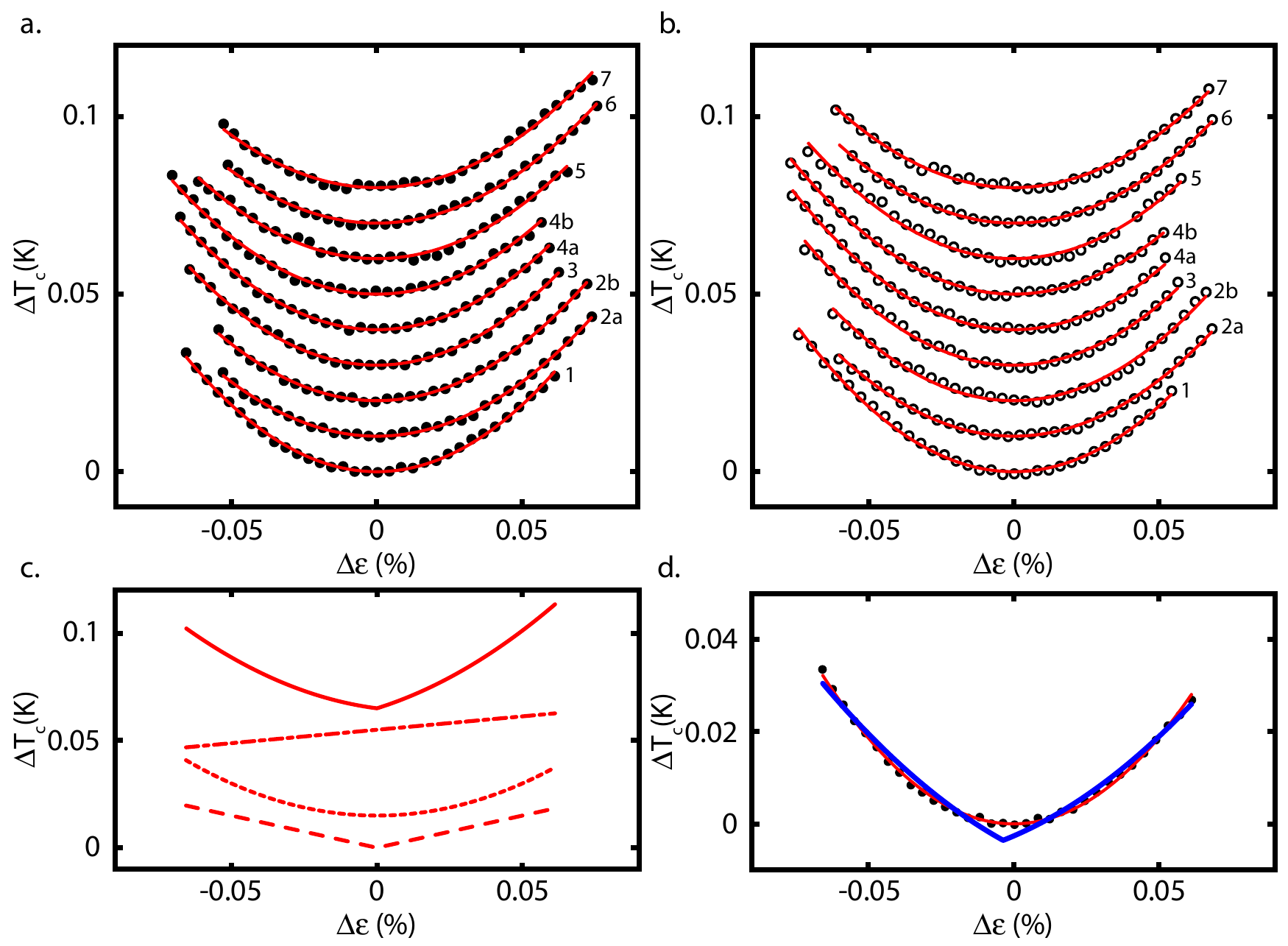}
\caption{\label{fits}Fitting individual curves of $T_c$ vs. strain. (a) Increasing (compressive to tensile) strain sweeps (black points), labeled at right by position, and quadratic fits (red); the origin is the minimum measured $T_c$ for each curve and the curves are offset for clarity. There is quantitative agreement between the data and fit without a cusp term. (b) The same for decreasing (tensile to compressive) strain sweeps (data are open circles). (c) Plot showing expected magnitudes of individual components, from bottom to top: 300 mK/\% cusp  (dashed); 6 K/\%$^2$ quadratic (dotted); 125 mK/\% linear (dash-dot); combined strain dependence (solid); (d) A single strain sweep (black points) from (a). The red line is a pure quadratic fit. The blue line is a fit including a cusp, with the cusp magnitude fixed at the previously calculated value $\gamma = 300$ mK/\%;\cite{Steppke_SRO} it clearly deviates from the data at low strains.} 
\end{figure*}

We first fit the data to a pure quadratic model, $T_c(\epsilon) = \alpha(\epsilon-\epsilon_0)^2 + T_{c,min}$, taking each strain sweep separately. Figure~\ref{fits}(a) and (b) show the $T_c$ vs. strain curves with the quadratic fits after shifting horizontally by $\epsilon_0$ and vertically by $T_{c,min}$. The fits are in excellent quantitative agreement with the data, even without a cusp ($|\epsilon|$) term. The extracted fit parameters are given in Table~\ref{fitpars}.  The average value of $\alpha$ over all of the fits is 6.47 K/\%$^2$, in good agreement with the values obtained from previous measurements over a wider strain range. Agreement between the data and this quadratic dependence persists to the lowest measured values of strain. While Ref.~\onlinecite{Hicks_SRO} reported an anomalous flattening in the strain dependence of $T_c$ at low strains, the present study shows that this was most likely an effect of inhomogeneity of the type that we observe directly here. 

Although there is no visually apparent cusp term in the data, we can explore the possible presence of a cusp term through fitting. We reference our expression to the pure quadratic fit by writing $\Delta \epsilon = \epsilon - \epsilon_0$ and $\Delta T_c = T_c - T_{c,min}$, with $\epsilon_0$ and $T_{c,min}$ the values obtained from the pure quadratic fit for each sweep. We then have $\Delta T_c(\Delta\epsilon) = \alpha(\Delta\epsilon-\epsilon_0^\prime)^2 + \beta(\Delta\epsilon-\epsilon_0^\prime) + \gamma|\Delta\epsilon-\epsilon_0^\prime| + dT_{c,min}$, where $\epsilon_0^\prime$ is the location of the cusp relative to $\epsilon_0$. In Ref.~\onlinecite{Steppke_SRO}, an expected cusp magnitude of $\gamma = 300$ mK/\% was calculated for $p_x \pm ip_y$ superconductivity by a renormalization group method. As a visual guide, we show in Fig.~\ref{fits}(c) the expected terms in $T_c(\Delta \epsilon)$ at their expected magnitudes, including a 300 mK/\% cusp. In Fig.~\ref{fits}(d), we show trace 1 from Fig.~\ref{fits}(a) with a fit including the cusp term with $\gamma$ held fixed at 300 mK/\%. The fit clearly deviates from the data; furthermore, when this fixed cusp term is included in fitting all of the curves, the average of the fitted values of $\alpha$ is reduced to 2.30 K/$\%^2$, much lower than the values obtained in previous measurements. We therefore conclude that a cusp term of the expected magnitude, $\sim$300 mK/\%, is inconsistent with our data.

To refine the upper bound on the cusp term, we fit with the following free parameters: $\alpha$, $\beta$, $\gamma$, $\epsilon_0^\prime$, and $dT_{c,min}$ (fits not shown). The 	quadratic ($\alpha$), linear ($\beta$), and cusp location ($\epsilon_0^\prime$) parameters are constrained to be positive. In Table~\ref{fitpars2}, we report the mean and 95\% confidence interval for each  parameter, obtained by bootstrapping. The fitted values for $\alpha$, $\beta$, and $\epsilon_0^\prime$ are in line with with expectations from previous measurements. The variability of the fitted values is amplified by the non-orthogonality of the parameters, especially the anti-correlation between the quadratic and cusp terms, which are both symmetric about $\epsilon_0^\prime$. Nevertheless, the fitted values from individual sweeps do not all agree with each other within their confidence intervals, indicating that there is likely a systematic variability from sweep to sweep. The most likely origin is slipping of the sample in the epoxy. Minor slipping is consistent with the observation that there is a notable difference between the average values of the best-fit cusp magnitudes $\gamma$ for the two sweep directions reported in Table~\ref{fitpars2}: $-5~(-92~114)$ mK/\% for increasing sweeps and $69~(-4~171)$ mK/\% for decreasing sweeps. Here, the mean and 95\% confidence interval for each were obtained from a combined distribution of the bootstrap iterations from all of the sweeps in each sweep direction.

A systematic, sweep-independent distortion of the applied strain that nearly cancels (and thereby hides) a larger cusp is possible in principle, but it seems more likely that the total systematic error in each sweep is comparable to the variation between sweeps. A cusp magnitude of 214 mK/\% is excluded at 95\% confidence in all sweeps individually and could be taken as an upper bound. Another estimate of the upper bound could be taken from the mean and confidence interval of $\gamma$ extracted from the combined distribution of all bootstrapped iterations for all sweeps, $32~(-81~157)$ mK/\%. We conclude that if any cusp is present, it is likely smaller than $\gamma<$ $\sim$150 mK/\%. 

\section{\label{sec:DiscConc}Discussion and Conclusion}
Anisotropic strain has already shown significant utility as a symmetry-breaking field for the study of collective electronic states including superconductivity\cite{Steppke_SRO,Barber_SRO,Palmstrom_Ba122,Kissikov_Ba122} and magnetism.\cite{Brodsky_327}  We have successfully integrated an apparatus for the \textit{in situ} application of uniaxial pressure with cryogenic scanning SQUID microscopy.  Using this setup, we have shown that the strongly quadratic response of the superconducting transition temperature of Sr$_2$RuO$_4$ to the application of uniaxial pressure, suggested by ultrasonic attenuation measurements\cite{Matsui_ultrasound, *Matsui_ultrasound2} and previously demonstrated by bulk ac susceptibility measurements in the presence of applied uniaxial pressure,\cite{Hicks_SRO} persists even to the lowest strains, where $T_c$ is within a few millikelvin of its minimum value. Our measurements indicate that an apparently flatter functional form at the lowest strain values in previous bulk measurements was therefore likely an effect of inhomogeneity.

Furthermore, we rule out the existence of a cusp at the level of 300 mK/\% that was recently estimated.\cite{Steppke_SRO}   In principle, a finite cusp at this level could be obscured by the effects of thermal fluctuations,\cite{Berg_fluctuations} but the effect of such fluctuations on the superconducting transition is small, as indicated by the linear dependence on temperature of the diamagnetic susceptibility near $T_c$.  

We have shown that using a local probe can greatly improve the sensitivity of $T_c$ measurements.  Future measurements using a more robust epoxy joint will enable more precise measurement of the strain as well as measurement of the superconducting transition of Sr$_2$RuO$_4$ over a broader range of strain.  These improvements should provide a tighter bound on the size of the cusp and will allow a determination of whether there is fine structure in the evolution of $T_c$ across the likely van Hove singularity,\cite{Steppke_SRO,Barber_SRO} while measurements of the temperature dependence of the penetration depth using scanning SQUID microscopy can investigate the possibility of a change in order parameter at finite strain.

\begin{acknowledgments}
We thank Hilary Noad for experimental assistance and feedback on the manuscript, and Ian Fisher, Wen Huang, Steven Kivelson, Johanna Palmstrom, and Thomas Scaffidi for helpful discussions.  This  work  was  supported  by the Department of Energy, Office of Science, Basic Energy Sciences, Materials Sciences and Engineering Division, under Contract No. DE-AC02-76SF00515.
\end{acknowledgments}

\bibliography{sro} 

\begin{table}
\caption{\label{fitpars}Extracted best fit parameters for pure quadratic fits to increasing [Fig.~\ref{fits}(a)] and decreasing [Fig.~\ref{fits}(b)] strain sweeps: quadratic coefficients ($\alpha$) in K/\%$^2$, offset strain at $T_{c,min}$ ($\epsilon_0$) in \%, and $T_{c,min}$ in K.}
\begin{ruledtabular}
\begin{tabular}{|l|llll|}
Point & \multicolumn{3}{c}{Increasing (Compressive to Tensile)}&\\
\#& $\alpha$ (K/\%$^2$)  &$\epsilon_0$ (\%) & $T_{c,min}$ (K)&\\
\hline
1&7.50&-0.022&1.51&\\
2a&6.09&-0.035&1.46&\\
2b&6.36&-0.033&1.46&\\
3&6.63&-0.024&1.48&\\
4a&6.73&-0.021&1.47&\\
4b&6.45&-0.019&1.47&\\
5&6.05&-0.027&1.48&\\
6&5.86&-0.080&1.51&\\
7&5.89&-0.077&1.48&\\
\hline
Point & \multicolumn{3}{c}{Decreasing (Tensile to Compressive)}&\\
\#& $\alpha$ (K/\%$^2$)  &$\epsilon_0$ (\%) & $T_{c,min}$ (K)&\\
\hline
1&7.35&-0.019&1.51&\\
2a&6.31&-0.033&1.46&\\
2b&6.69&-0.031&1.46&\\
3&6.76&-0.021&1.47&\\
4a&6.76&-0.017&1.47&\\
4b&6.46&-0.016&1.47&\\
5&6.51&-0.022&1.48&\\
6&6.12&-0.077&1.51&\\
7&5.99&-0.074&1.48&
\end{tabular}
\end{ruledtabular}
\end{table}

\begin{table*}
\caption{\label{fitpars2}Mean values and 95\% confidence intervals for fit parameters obtained by bootstrapping fits containing a cusp term (fits not shown):  quadratic coefficients ($\alpha$) in K/\%$^2$; linear coefficients ($\beta$) in mK/\%; cusp magnitudes ($\gamma$) in mK/\%; cusp locations relative to the $T_c$ minimum in quadratic-only fits ($\epsilon_0^\prime$) in \%, and shifts of $T_{c,min}$ relative to $T_{c,min}$ in quadratic-only fits ($dT_{c,min}$) in mK.}
\begin{ruledtabular}
\begin{tabular}{|l|llllll|}
Point & \multicolumn{5}{c}{Increasing (Compressive to Tensile)}&\\
\#&$\alpha$ (K/\%$^2$)& $\beta$ (mK/\%)& $\gamma$ (mK/\%)& $\epsilon_0^\prime$ (\%)&$dT_{c,min}$ (mK)&\\
\hline
1 & $7.89~ (7.30~ 8.39)$ & $130~ (83~ 208)$ & $-30~ (-65~ 5)$ & $0.008~ (0.006~ 0.013)$ & $1.0~ (0.3~ 2.2)$&\\
2a & $6.55~ (6.02~ 6.99)$ & $131~ (87~ 179)$ & $-32~ (-61~ 2)$ & $0.011~ (0.007~ 0.015)$ & $1.1~ (0.5~ 1.6)$&\\
2b & $6.69~ (6.10~ 7.27)$ & $114~ (65~ 174)$ & $-24~ (-59~ 12)$ & $0.009~ (0.005~ 0.014)$ & $0.8~ (0.1~ 1.6)$&\\
3 & $6.34~ (5.93~ 6.78)$ & $114~ (75~ 176)$ & $22~ (-7~ 51)$ & $0.009~ (0.006~ 0.013)$ & $0.2~ (-0.4~ 0.9)$&\\
4a & $7.08~ (6.63~ 7.44)$ & $121~ (84~ 165)$ & $-27~ (-55~ 1)$ & $0.009~ (0.006~ 0.012)$ & $0.9~ (0.4~ 1.6)$&\\
4b & $7.38~ (6.70~ 7.89)$ & $117~ (58~ 171)$ & $-75~ (-114~ -28)$ & $0.009~ (0.004~ 0.013)$ & $1.6~ (0.9~ 2.4)$&\\
5 & $4.81~ (3.95~ 5.44)$ & $180~ (100~ 265)$ & $97~ (53~ 152)$ & $0.015~ (0.009~ 0.021)$ & $0.1~ (-1.5~ 1.5)$&\\
6 & $5.24~ (4.62~ 5.93)$ & $54~ (3~ 95)$ & $45~ (6~ 92)$ & $0.004~ (0.000~ 0.008)$ & $-0.4~ (-1.3~ 0.2)$&\\
7 & $6.15~ (4.79~ 7.60)$ & $125~ (62~ 198)$ & $-17~ (-98~ 69)$ & $0.010~ (0.005~ 0.017)$ & $0.9~ (-0.5~ 1.9)$&\\
\hline
Point & \multicolumn{5}{c}{Decreasing (Tensile to Compressive)}&\\
\#&$\alpha$ (K/\%$^2$)& $\beta$ (mK/\%)& $\gamma$ (mK/\%)& $\epsilon_0^\prime$ (\%)&$dT_{c,0}$ (mK)&\\
\hline
1 & $6.42~ (5.98~ 6.94)$ & $130~ (81~ 177)$ & $78~ (45~ 113)$ & $0.009~ (0.006~ 0.012)$ & $-0.6~ (-1.3~ 0.2)$&\\
2a & $5.98~ (5.54~ 6.71)$ & $206~ (136~ 313)$ & $26~ (-18~ 62)$ & $0.017~ (0.011~ 0.025)$ & $1.4~ (0.5~ 3.1)$&\\
2b & $5.50~ (5.07~ 5.88)$ & $362~ (176~ 500)$ & $114~ (70~ 153)$ & $0.027~ (0.014~ 0.036)$ & $3.3~ (0.3~ 6.7)$&\\
3 & $5.67~ (5.05~ 6.38)$ & $132~ (70~ 222)$ & $90~ (47~ 136)$ & $0.010~ (0.006~ 0.016)$ & $-0.6~ (-1.2~ 0.3)$&\\
4a & $6.28~ (5.97~ 6.66)$ & $242~ (152~ 424)$ & $51~ (16~ 92)$ & $0.018~ (0.012~ 0.029)$ & $1.4~ (0.0~ 4.4)$&\\
4b & $6.33~ (5.89~ 6.82)$ & $118~ (58~ 182)$ & $13~ (-20~ 50)$ & $0.009~ (0.004~ 0.014)$ & $0.3~ (-0.8~ 1.3)$&\\
5 & $4.68~ (3.86~ 5.57)$ & $149~ (85~ 206)$ & $152~ (99~ 214)$ & $0.012~ (0.007~ 0.016)$ & $-1.3~ (-2.7~ -0.4)$&\\
6 & $5.77~ (5.40~ 6.16)$ & $175~ (145~ 216)$ & $23~ (-3~ 45)$ & $0.014~ (0.012~ 0.018)$ & $1.0~ (0.5~ 1.6)$&\\
7 & $5.11~ (4.52~ 5.65)$ & $216~ (161~ 299)$ & $68~ (29~ 111)$ & $0.018~ (0.014~ 0.025)$ & $1.1~ (0.3~ 2.8)$&
\end{tabular}
\end{ruledtabular}
\end{table*}
\end{document}